\documentclass[pra,amsmath,superscriptaddress,twocolumn,showpacs]{revtex4-1}
\usepackage{bm}
\usepackage{amssymb}
\usepackage{amsmath}
\usepackage{colordvi}
\usepackage{graphicx}
\usepackage{color}
\usepackage{hyperref}

\def\spin{{\mbox{\boldmath{${\sigma}$}}}}

\newcommand{\ov}[1]{\overline{{#1}}}
\newcommand{\be}{\begin{equation}}
\newcommand{\ee}{\end{equation}}
\newcommand{\bea}{\begin{eqnarray}}
\newcommand{\eea}{\end{eqnarray}}

\def\nn{\nonumber}

\def\ket#1{\vert #1 \rangle}
\def\bra#1{\langle #1 \vert}
\def\grad {\mbox{\boldmath$\nabla$\unboldmath}}

\begin{document}



\title{Tunable topological Weyl semimetal from simple cubic lattices
  with staggered fluxes}

\author{Jian-Hua Jiang}
\email{jianhua.jiang.phys@gmail.com}
\affiliation{Department of Condensed Matter Physics, Weizmann Institute of
  Science, Rehovot 76100, Israel}

\date{\today}

\begin{abstract}
Three-dimensional Weyl fermions are found to emerge from simple cubic
lattices with staggered fluxes. The mechanism is to gap the quadratic
band touching by time-reversal-symmetry-breaking hoppings. The system
exhibits rich phase diagrams where the number of Weyl fermions and
their topological charge are tunable via the plaquette fluxes. The Weyl
semimetal state is shown to be the intermediate phase between
non-topological semimetal and quantum anomalous
Hall insulator. The transitions between those phases can be understood
through the evolution of the Weyl points as Berry flux insertion
processes. As the Weyl points move and split (or merge) through tuning
the plaquette fluxes, the Fermi arcs and surface states undergo
significant manipulation. We also propose a possible scheme to
realize the model in ultracold fermions in optical lattices with
artificial gauge fields.
\end{abstract}

\pacs{71.10.Fd, 03.75.Ss, 05.30.Fk, 03.65.Vf}

\maketitle

\section{Introduction} 
Massless Dirac fermions possess chiral symmetry and can be classified
by their chirality $\gamma$. For example, in three-dimension (3D), if
the Hamiltonian is $H=v_F\spin\cdot{\bf q}$ where $\spin$ is the Pauli
matrices vector and ${\bf q}$ is the wavevector, then the chirality is
$\gamma={\rm sgn}(v_F)$. Such fermions (also called chiral or Weyl
fermions) possess several peculiar behaviors such as Adler-Bell-Jackiw
anomaly\cite{anomaly}. Chiral fermions in two-dimension (2D) have been
found in real condensed matter systems since the discovery of
graphene\cite{graphene1}. Many novel electronic properties of
graphene\cite{graphene2}: Klein tunneling, the peculiar integer
quantum Hall effect, the transport properties such as the conductivity
minimum, the weak (anti)localization, and edge states, originate from
the chiral fermion nature. Recent studies also found the realizations
of 2D chiral fermions in ultracold atomic gases optical
lattices\cite{2d}. A generic route to chiral fermions is to search
systems with two-band touching\cite{note0}. Around the band touching
node, the Hamiltonian can generally be written as $H={\bf h}({\bf
  k})\cdot\spin$ where Pauli matrices $\spin$ act on the space spanned
by the two bands and $|{\bf h}|\to 0$ at the node. For example, in
graphene, around one of the node ${\bf K}$, ${\bf h}({\bf
  k})=v_F({\bf k}-{\bf K})$. Thus chiral fermions emerge as
topological defects (vortices) in ${\bf k}$-space. They are generally
classified by their vortex winding number $N_w$. In the special cases
when $N_w=\pm 1$, the winding number gives the chirality $\gamma=N_w$.
In general cases, $N_w$ can be any integer. The total
winding number in the system must be conserved under adiabatic
transformation. The winding number can be defined by the Berry phase
carried by the node, $N_w = \frac{1}{\pi}\oint_{{\Gamma}} d{\bf
  k}\cdot \langle \Psi({\bf k})|i\grad_{\bf k}|\Psi({\bf k})\rangle$,
where ${\Gamma}$ is a contour enclosing the node and $\Psi({\bf k})$
is the single-valued and continuous wavefunction of the eigenstates
with $H\ket{\Psi({\bf k})}=\pm |{\bf h}({\bf k})|\ket{\Psi({\bf
    k})}$. By breaking the time-reversal symmetry, such two-band
touching can be gapped, leading to a quantum anomalous Hall (QAH)
insulator with Chern number $C=\pm \frac{1}{2}N_w$\cite{pairing}.

3D Weyl fermions are more robust: they can not even be gapped by
time-reversal symmetry breaking. In fact they can only be annihilated
in pairs with the total winding number conserved under adiabatic
transformations. The winding number of 3D Weyl fermion is defined
as\cite{Volovik} $N_w=\frac{1}{8\pi}\varepsilon^{\nu\delta\rho}\oint_{\cal S} 
dS^\rho {\hat {\bf n}}\cdot\Big(\partial_{k_\nu}{\hat {\bf
    n}}\times \partial_{k_\delta}{\hat {\bf n}}\Big)$ 
($\nu,\delta,\rho=x,y,z$ and $\varepsilon$ being the Levi-Civita tensor)
where ${\cal S}$ is a surface enclosing the band touching node (Weyl
point), $dS^\rho$ is the surface area elements along $\rho$ direction,
and ${\hat {\bf n}}={\bf h}/|{\bf h}|$. 3D Weyl fermions are monopoles
of the Berry-phase gauge fields where the monopole charge is the
topological charge $N_w$\cite{Volovik}. As a consequence, there is a
step change in the Hall conductivity, e.g., $\sigma_{xy}(k_z)$  [and
the Chern number $C_{xy}(k_z)=\frac{h}{e^2}\sigma_{xy}(k_z)$] as
function of $k_z$,
\be
\sigma_{xy}(k_z)={\rm sgn}(k_z-k_z^c) \frac{N_w}{2} \frac{e^2}{h} +
... ,
\label{dnc}
\ee
due to the monopole at $k_z^c$, while ... denotes other contributions to
the Hall conductivity (Chern number). According to the Nielsen-Ninomiya
theorem\cite{fdt}, Weyl points must appear in pairs with opposite $N_w$
in a lattice system.

Weyl semimetal (system with 3D Weyl fermions) possesses very special
properties such as chiral surface states with open Fermi surfaces 
(Fermi arcs) terminating at the projection of the Weyl points and
thickness dependent quantized anomalous Hall conductivity in thin
films, which are first found in the studies of
$^3$He-A\cite{Volovik,early}. After that Murakami showed that a Weyl
semimetal phase can appear as intermediate phase between normal
insulator and topological insulator\cite{muraki}. Recent studies
demonstrate that Weyl semimetal can also be realized in other condense
matter systems:\cite{phys} some pyrochlore iridates (such as
Y$_2$Ir$_2$O$_7$)\cite{Wan}, superlattices made of topological insulator and
non-topological insulator thin films with broken
time-reversal\cite{Burkov1,Burkov2} or inversion\cite{balents}
symmetry, the ferromagnetic compound HgCr$_2$Se$_4$\cite{iop,Fang},
and bulk magnetically doped Bi$_2$Se$_3$\cite{cho}. It is also found
that in some situations the number and type of the Weyl points are
determined and protected by the lattice
symmetries\cite{symmetry,Fang}. A topological nodal semimetals where
the bulk spectrum exists nodal lines are also proposed and studied in
Ref.~\cite{Burkov2}. There are also some lattice models where Weyl
fermions are found\cite{lattice}.

In this work we show that Weyl fermions can emerge from simple cubic
lattices with staggered fluxes through plaquettes
[see Fig.~1]. Differing from previous studies where the Weyl 
semimetal phases emerge due to inverted bands with spin-orbit
interactions\cite{iop,Fang,muraki,balents} or gaping Dirac cones by
various mass terms\cite{Burkov1,Burkov2,cho}, here the mechanism is to
gap the quadratic band touching via time-reversal-symmetry-breaking
hoppings due to the staggered fluxes. A direct distinction is that
such scenario does not need to invoke spin degeneracy breaking (i.e.,
spin-orbit coupling). The system exhibits rich phase diagrams where
the number of Weyl fermions and their topological charge are tunable
via the fluxes per plaquettes. The Weyl semimetal state is
demonstrated as the intermediate phase between non-topological
semimetal and quantum anomalous Hall insulator. The transitions
between those phases can be understood via the evolution of the Weyl
points as Berry flux insertion processes [see Sec.~IV and Summary
section]. As the Weyl points move and split (or merge) through tuning
the plaquette fluxes, the Fermi arcs and surface states undergo
significant change, because the Fermi arcs act as Dirac strings which
have to connect the monopoles with opposite charges. Finally, we
propose a possible scheme to realize such model in ultracold fermionic
gases in optical lattices with artificial gauge fields.

\section{Lattice and Hamiltonian} 
We consider a simple cubic lattice
system, which can be viewed as stacking of layers of 2D lattices with
checkerboard-patterned staggered fluxes [see
Fig.~1]\cite{Sun2,Sun,our,ourqbt}. The 2D checkerboard lattice is
designed in such a way that the hopping between the nearest-neighbored
A-type sites along the $x$ and $y$ directions are $t_x$ and $t_y$
respectively, whereas those for B-type sites are $t_y$ and $t_x$
respectively [see Fig.~1(a)]. The hopping from A-type site to the
nearest B-type ones are $t_2e^{\pm i\phi_1}$ as indicated in Fig.~1(a). The
flux per plaquette is $\pm \Phi_1=\pm 4\phi_1$ [Fig.~1(a)]. Recently Sun
{\sl et al.} proposed a realistic optical lattice system to realize
such model\cite{Sun2}. Besides it can be realized in ultracold
fermions in optical lattices and in condensed matter systems with
artificial\cite{exp,the1,the2,the3,design} or ``emergent''\cite{emerg}
gauge fields. The Hamiltonian for each layer
is\cite{Sun2,Sun,our,ourqbt} $H_{\rm 2D} = h_0({\bf k})\sigma_0 + {\bf 
  h}_1({{\bf k}}) \cdot\spin$ where $\spin$ is the Pauli matrix
vector acting on A/B (pseudo-spin up/down) site space and
$\sigma_0$ is the $2\times 2$ identity matrix. As it is no need to
invoke true-spin splitting, we keep the true-spin states as 
degenerate. $h_0({\bf k})=2t_0(\cos k_x +\cos k_y)$,  $h_{1z}({{\bf
    k}})=2t_1(\cos k_x - \cos k_y)$, $h_{1x}({{\bf 
    k}})=4t_2\cos\phi_1\cos\frac{k_x}{2} \cos\frac{k_y}{2}$, and
$h_{1y}({\bf k})=4t_2\sin\phi_1\sin\frac{k_x}{2}\sin\frac{k_y}{2}$ where
$t_0=(t_x+t_y)/2$, $t_1=(t_x-t_y)/2$ [Note that throughout this
paper, we set the lattice constant $a=1$. All the wavevectors are then
in the units of $1/a=1$.]. The two bands touches
quadratically at $(\pi,\pi)$ when $\sin\phi_1=0$. This band touching
is nontrivial as it carries a nonzero winding number $N_w=2{\rm
  sgn}(t_1t_2\cos\phi_1)=\pm 2$. At finite $\sin\phi_1$
(broken time-reversal symmetry) the quadratic band touching is gapped
and the system becomes a QAH insulator with Chern number
$C=\frac{1}{2}{\rm sgn}(t_2\sin\phi_1)N_w$\cite{Sun,our}.

\begin{figure}[htb]
\includegraphics[height=3.3cm]{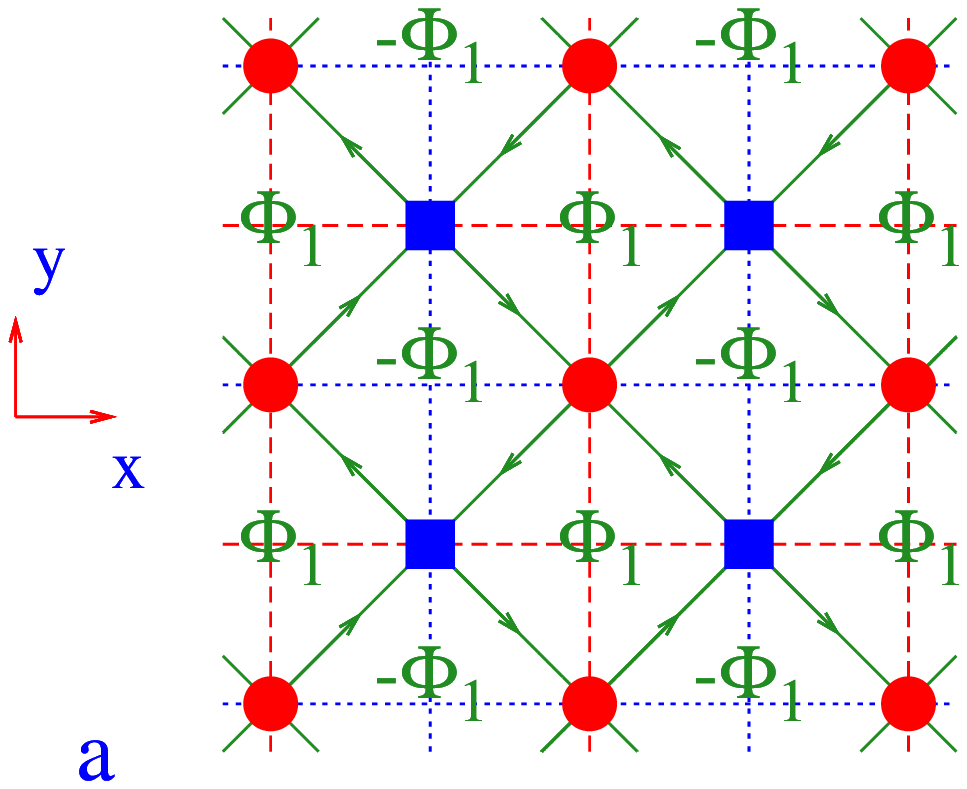}\includegraphics[height=3.3cm]{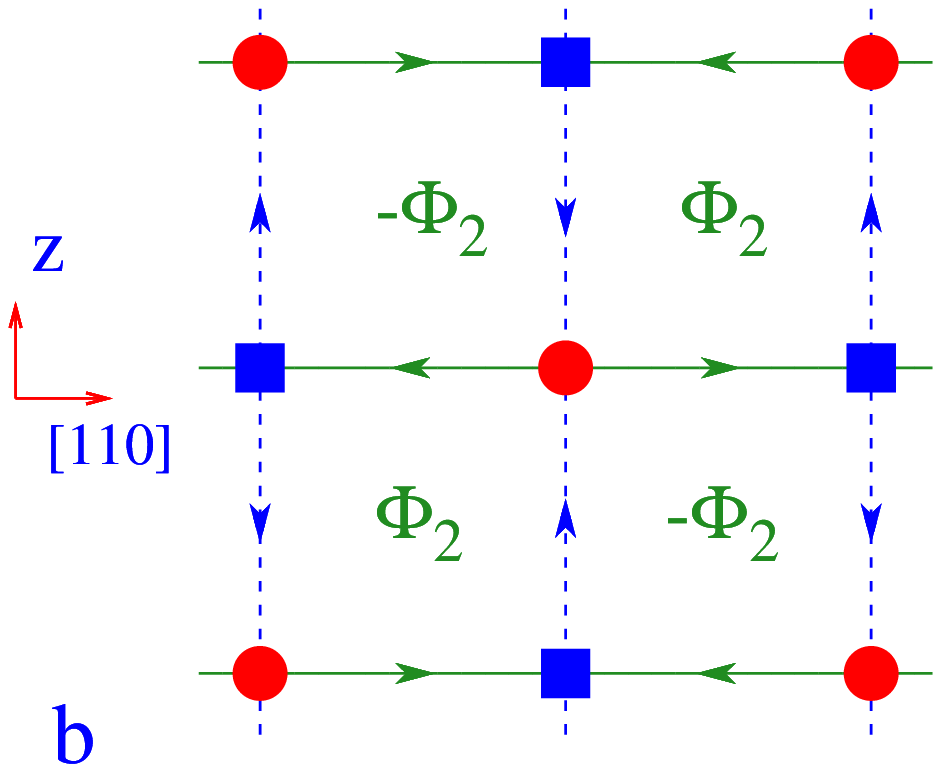}
\includegraphics[height=3.8cm]{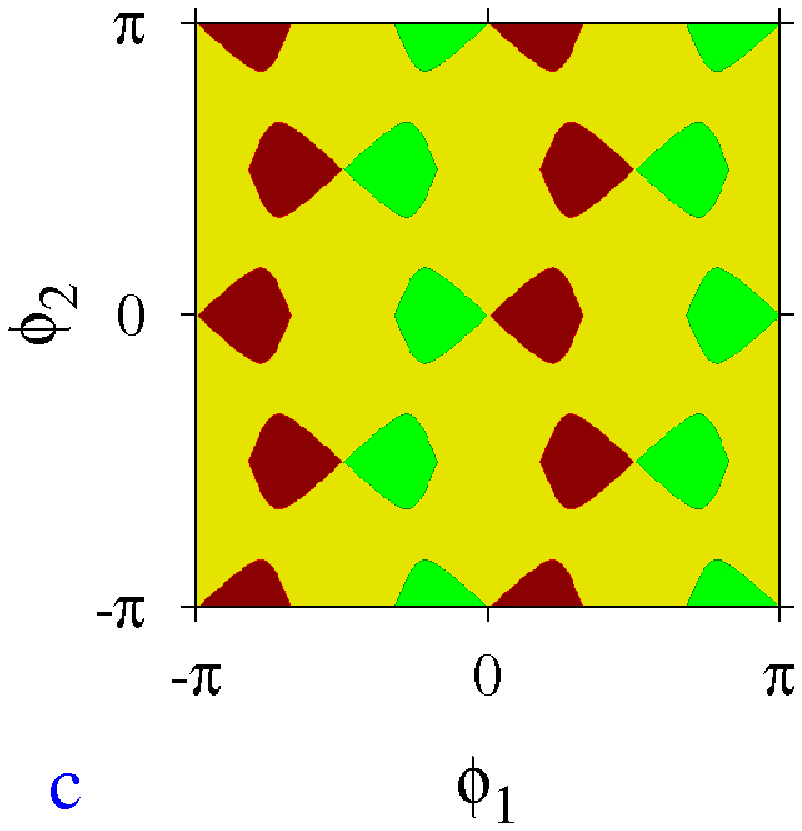}\includegraphics[height=3.8cm]{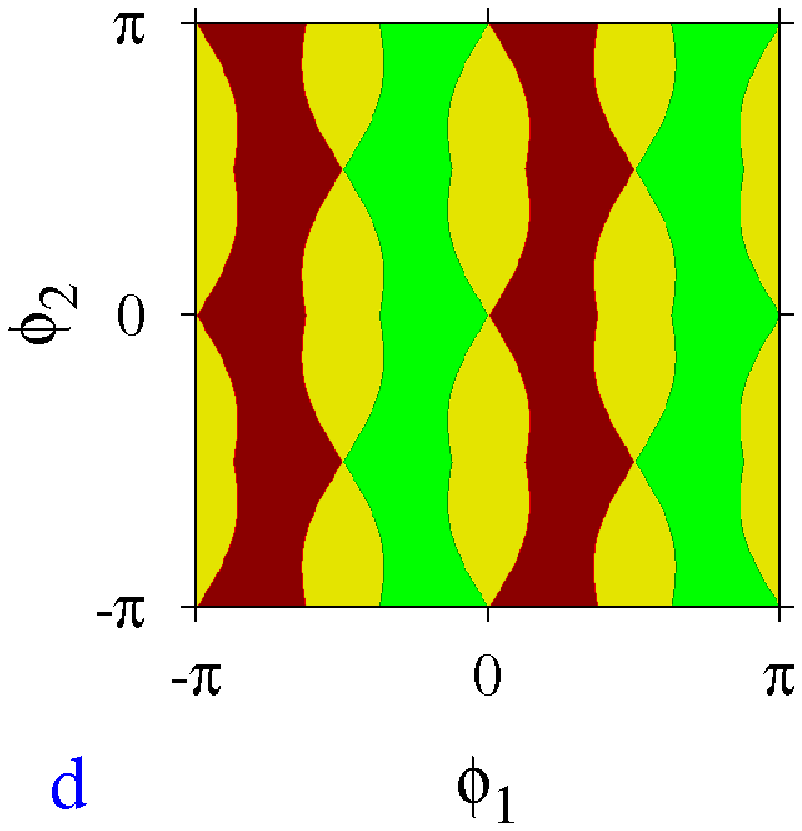}
\caption{(Color online) Lattice structure in (a) (001) plane in an odd
  layer and (b)($1\ov{1}0$) plane. Red-dots and blue-squares represent type A
  and B sites respectively. Hoppings along (opposite) the
  green-solid and blue-dotted arrows are $t_2e^{-i\phi_1}$ and
  $t_{\perp}e^{-i\phi_2}$ ($t_2e^{i\phi_1}$ and
  $t_{\perp}e^{i\phi_2}$) separately. Those along red-dashed and
  blue-dotted lines are $t_y$ and $t_x$ respectively. $\pm \Phi_1$
  and $\pm \Phi_2$ are the plaquette fluxes. Phase diagram at
  (c) $|t_2/t_\perp|=0.95$ and (d) $|t_2/t_\perp|=1.3$. Yellow region:
  Weyl semimetal; green region: QAH state with Chern number
  $C_{xy}(k_z)= -2{\rm sgn}(t_1t_2t_{\perp})$ for all $k_z$; brown
  region: QAH state with $C_{xy}(k_z)= 2{\rm sgn}(t_1t_2t_{\perp})$
  for all $k_z$.}
\label{fig1}
\end{figure}

The 3D lattice is a stacking of the 2D layers in such a way that
different types of sites are on top of each other [see Fig.~1(b)].
The hopping between those sites is $t_{\perp}e^{\pm i\phi_2}$. The
staggered flux per plaquette in the $(1\ov{1}0)$ plane is $\pm \Phi_2=\pm
2(\phi_1+\phi_2)$ [in $(110)$ plane, the plaquette flux is $\pm
2(\phi_1-\phi_2)$]. The Hamiltonian of the system is then
\bea
H &=& [h_0({\bf k})\sigma_0+{\bf h}({\bf k})\cdot\spin]\tau_0 \nn \\
&&\mbox{} + g(k_z)(\cos\phi_2\sigma_x+\sin\phi_2\sigma_y) \tau_x 
\eea
Here $\sigma$ and $\tau$ matrices act on the A/B sites and
odd/even layers respectively. $g(k_z)=2t_{\perp}\cos \frac{k_z}{2}$. The
Hamiltonian can be block diagonalized directly, giving
\be
H_{\pm} = h_0({\bf k})\sigma_0 + {\bf h}_{\pm}({\bf k})\cdot\spin
\ee
in the two blocks where
\bea
h_{\pm x}({\bf k}) &=& 4t_2\cos\phi_1\cos\frac{k_x}{2} \cos\frac{k_y}{2} \pm
g(k_z)\cos\phi_2 ,\nn \\
h_{\pm y}({\bf k}) &=& 4t_2\sin\phi_1\sin\frac{k_x}{2}\sin\frac{k_y}{2} \pm
g(k_z)\sin\phi_2 ,\nn \\
h_{\pm z}({\bf k}) &=& 2t_1(\cos k_x - \cos k_y) .
\label{H0}
\eea
For the sake of easier formulation, we set the first Brillouin zone
as $k_x\in [0,2\pi]$, $k_y\in [0,2\pi]$ and $k_z\in [0,2\pi]$. Finally,
the system possesses a series inversion symmetries. It is invariant
under the following inversion transformations: (i) $z\to -z$, (ii)
$(x,y)\to (-x,-y)$, and (iii) $(x,y,z)\to (-x,-y,-z)$. At $h_0=0$, the
system also has particle-hole symmetry.

\section{Phase diagrams} 
The energy spectrum of the system is 
\be
E_{\alpha \pm}({\bf k}) = h_0({\bf k})\pm |{\bf h}_{\alpha}({\bf k})| 
\ee
with $\alpha=\pm$. The equations for the nodes (Weyl points) are
$h_{\alpha x}= h_{\alpha y}=h_{\alpha z}=0$. One finds that away from $\phi_1$, or
$\phi_2=0, \pm \pi/2, \pi$ [see Fig.~1 and discussions below], there
are four Weyl points ${\bf K}_{+\beta}=(\beta k_x^c, \beta k_y^c,
k_z^c)$ ($\beta=\pm$) in the $+$ block and ${\bf K}_{-\beta}=(\beta
k_x^c,\beta k_y^c, 2\pi-k_z^c)$ in the $-$ block where
\bea
&& k_y^c = \pi - 2\arctan\sqrt{\left|\frac{\tan\phi_1}{\tan\phi_2}\right|} ,\nn\\
&& k_x^c = \eta k_y^c \quad {\rm with}\quad \eta={\rm
  sgn}\left(\frac{\tan\phi_1}{\tan\phi_2}\right) ,\nn \\
&& k_z^c = 2\arccos\left[\frac{-2\eta t_2\sin\phi_1}{t_{\perp}\sin\phi_2}
\left(1+\left|\frac{\tan\phi_1}{\tan\phi_2}\right|\right)^{-1}\right]
.
\label{node}
\eea
Note that the above equation holds only when
\be
\left |
\frac{2 t_2\sin\phi_1}{t_{\perp}\sin\phi_2}
\left(1+\left|\frac{\tan\phi_1}{\tan\phi_2}\right|\right)^{-1}
\right| \le 1 ,
\ee
which sets the phase boundaries of the Weyl semimetal.
It is required that $|h_0({\bf k})-h_0({\bf 
  K}_{\alpha\beta})|\le |{\bf h}_{\alpha}({\bf k})|$ so that the
system is an insulator wherever away from the nodes ${\bf
  K}_{\alpha\beta}$ (the definition of semimetal). Given this, the
$h_0({\bf k})\sigma_0$ term is irrelevant for the physics to be
discussed and we hence take $h_0\equiv 0$ hereafter.

\begin{figure}[htb]
\includegraphics[height=3.55cm]{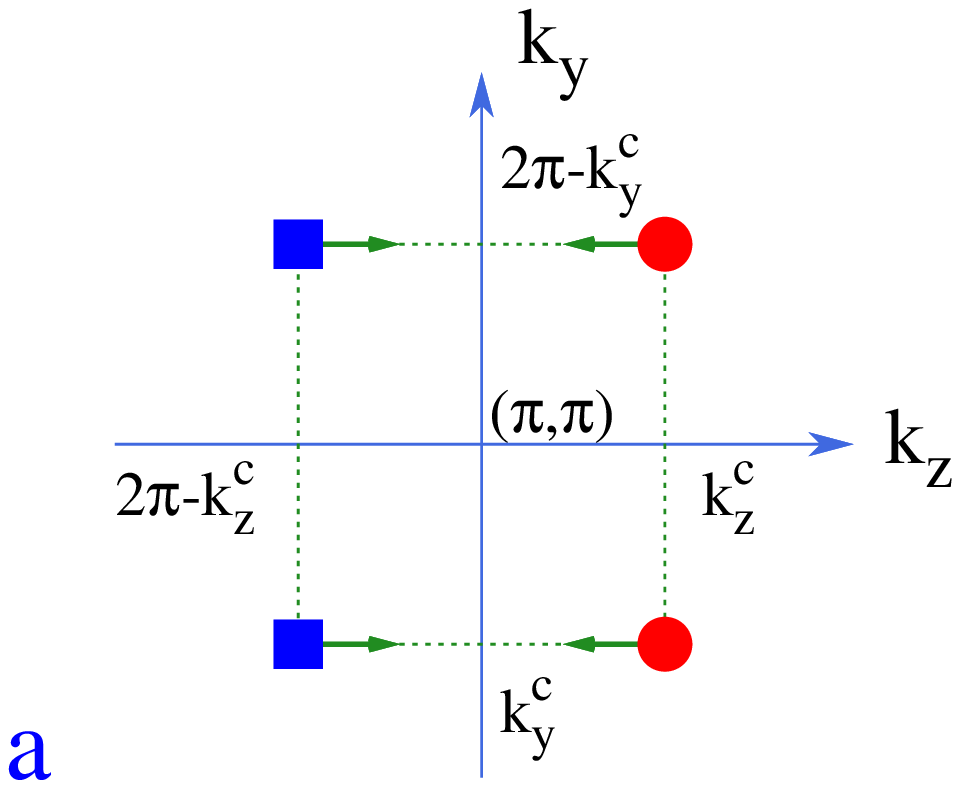}\includegraphics[height=3.4cm]{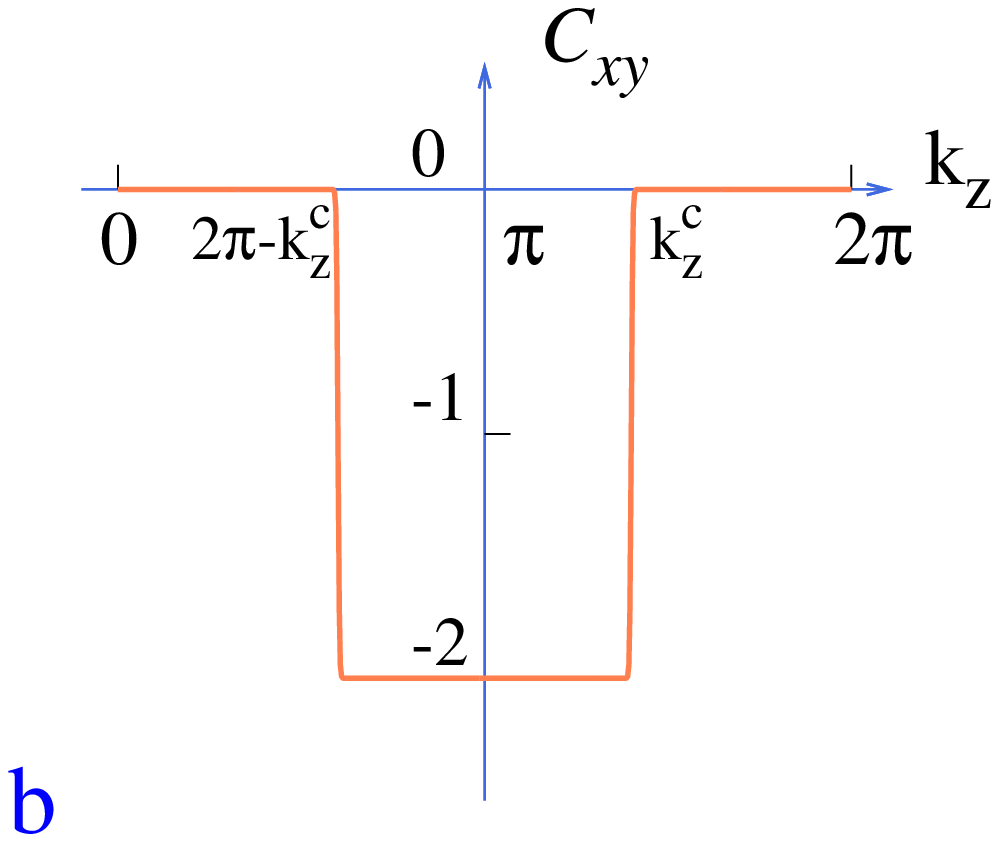}
\includegraphics[height=3.55cm]{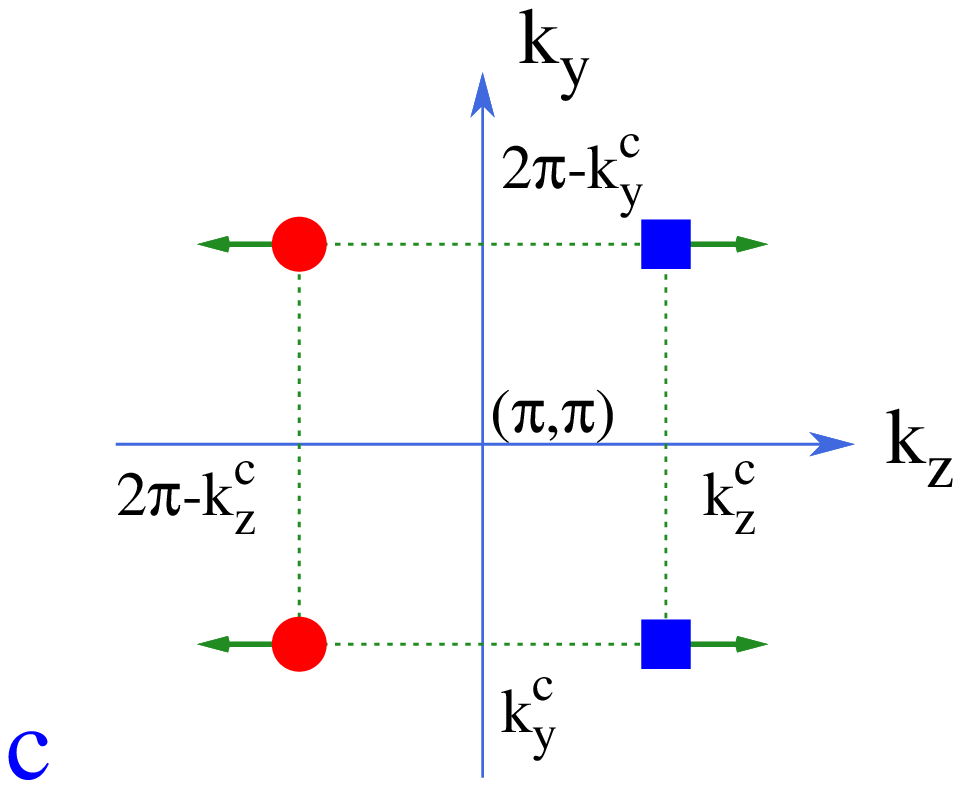}\includegraphics[height=3.4cm]{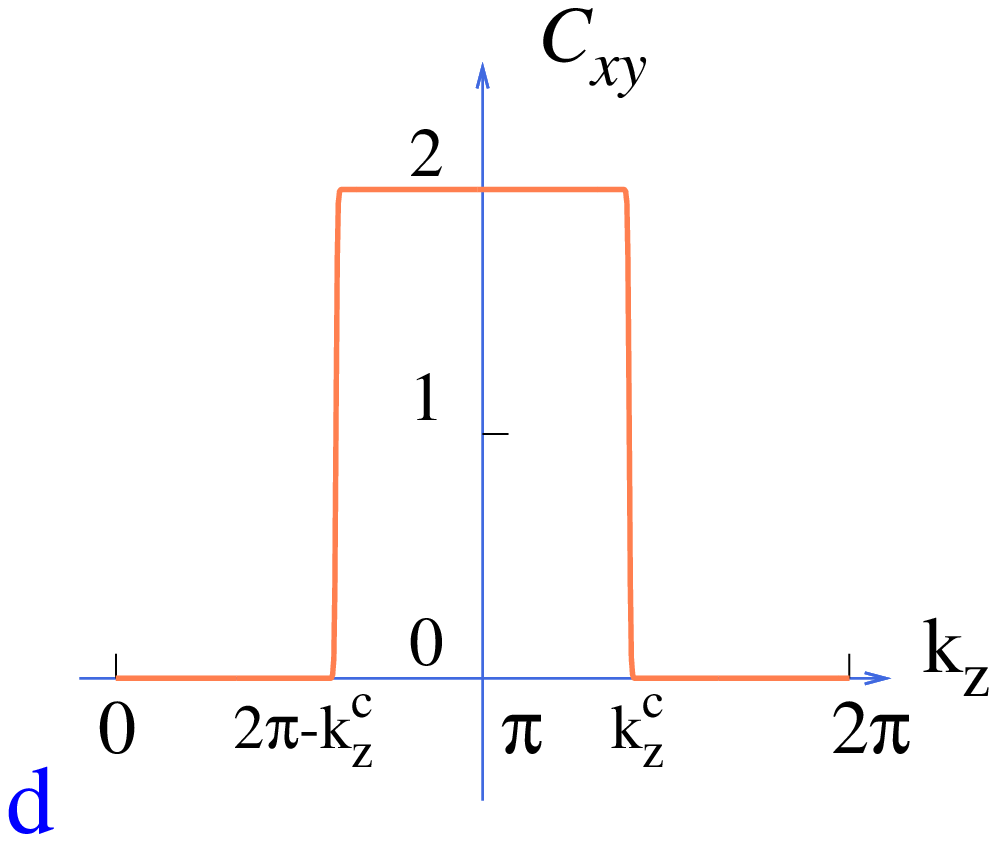}
\caption{(Color online) (a) and (c) Illustration of the Weyl points
  when there are four in $k_y$-$k_z$ plane. The arrows indicates the
  $z$-direction movement of the Weyl points as $\phi_1$
  increases. Blue-squares (red-dots) denote Weyl points with
  $N_w=-1$ ($N_w=1$). The parameters are $t_1=1$, $t_2=2$,
  $t_{\perp}=2$, and $\phi_2=-0.25\pi$. $\phi_1=-0.1\pi$ in (a), and
  $\phi_1=0.1\pi$ in (c). (b) and (d) Chern number $C_{xy}(k_z)$ as
  function of $k_z$ for the cases in (a) and (c) respectively.}
\end{figure}

In Fig.~1(c) and 1(d) we plot the phase diagram at two different
$|t_2/t_{\perp}|$. There are three phases: the Weyl semimetal,
the QAH, and the non-topological semimetal. The last one refers to
semimetals which have only nodes with $N_w=0$. Such nodes are
accidental band degeneracy points, which can be gapped by
infinitesimal band mixing without breaking any symmetries. The QAH
phase actually consists of two topologically distinct phases with
opposite Chern number [$C_{xy}(k_z) = \pm 2{\rm sgn}(t_1t_2t_{\perp})$
for all $k_z$]. Those two phases are separated by the Weyl semimetal
and non-topological semimetal phases. In fact, the Weyl semimetals
can be viewed as the intermediate phases between the non-topological
semimetal (or insulator) phase and the QAH phase. In the former the
Chern number is always zero, whereas in the latter it is always
nonzero. In Weyl semimetals, there are two regions in the Brillouin
zone: the Chern number is zero in one region while nonzero in
another. Specifically in the current model $C_{xy}(k_z)$ is nonzero
only when $k_z\in(k_z^{\rm min}, k_z^{\rm max})$ with $k_z^{\rm
  min}={\rm Min}(k_z^c, 2\pi-k_z^c)$ and $k_z^{\rm max}={\rm
  Max}(k_z^c, 2\pi-k_z^c)$. The QAH and non-topological semimetal
phases can be viewed as the limit $k_z^{\rm min}\to 0$ and $k_z^{\rm
  min}\to \pi$ respectively. There are actually two Weyl semimetal
phases in which the Chern number $C_{xy}(k_z)$ is opposite at
$k_z\in(k_z^{\rm min}, k_z^{\rm max})$. Those two states are the
intermediate states between the two QAH states (with opposite Chern
number) and the non-topological semimetal state separately. The
latter lies at the lines $\phi_1=0,\pm \pi/2$, and $\pi$ in the phase
diagrams, where $k_z^{\rm min}=k_z^{\rm  max}=\pi$. We will show in
next section that the evolution of the ground state and quantum phase
transitions between those phases can be understood via the evolution
of the Weyl points as Berry flux insertion processes.

We find that as $|t_2/t_{\perp}|$ increases, the area of the Weyl
semimetal phase in the phase diagram as functions of $\phi_1$ and
$\phi_2$ shrinks. At $|t_2/t_{\perp}|\to \infty$, such area becomes
zero. When $|t_2/t_{\perp}|\le 1/2$ the system is always in the Weyl
semimetal phase except at the special lines
$\phi_1=0,\pm\pi/2,\pi$. Hence the region of the Weyl semimetal phases
{\em can be tuned by the ratio $|t_2/t_{\perp}|$}. It is noted that
the phase diagram exhibits some angle-shaped structures around
the special points $(\phi_1,\phi_2)=(\phi_c, \pm \phi_c)$ with
$\phi_c=0, \pm \pi/2, \pi$ as well as $(0,\pi)$ and $(\pi,0)$ where
Eq.~(\ref{node}) becomes ill-defined. The structure around, say,
$(0,0)$, can be understood via the following analysis. For
$(\phi_1,\phi_2)$ close to $(0,0)$, along the line $\phi_2=\xi\phi_1$,
$k_z^c = 2\arccos[-2t_2/t_{\perp}(|\xi|+1)]$. Hence the Weyl semimetal
phase is at $|\xi|>|2t_2/t_{\perp}|-1$, which is angle-shaped. When
$|t_2/t_{\perp}|<1/2$, the system is in the Weyl semimetal phase for
all parameters $(\phi_1,\phi_2)$ around $(0,0)$. 

It is benefit to point out some special regions in the phase
diagram. First, when $\phi_2=\pm \pi/2$, there are only two
Weyl points: ${\bf K}_+=(\pi, \pi, k_z^c)$ in the 
$+$ block and ${\bf K}_-=(\pi,\pi,2\pi-k_z^c)$ in the $-$ block. Those
Weyl points are quadratic band touchings with winding number $N_w=\pm
2$. Similarly, when $\phi_2=0, \pi$, there are
two quadratic-band-touching Weyl points: ${\bf K}_+=(0, 0, k_z^c)$ in the
$+$ block and ${\bf K}_-=(0,0,2\pi-k_z^c)$ in the $-$ block.
Besides, as pointed out before at the lines $\phi_1=0,\pm\pi/2,\pi$
the system is a non-topological semimetal.

\section{Evolution of Weyl points} 
Away from the above regions, there are four Weyl points, around which
fermions are described by the Weyl Hamiltonian
\bea
H_{\alpha\beta}({\bf k}) = \spin\cdot \hat{v}_{\alpha\beta}\cdot {\bf q} + O(q^2)
\eea
where ${\bf q}={\bf k}-{\bf K}_{\alpha\beta}$. The velocity tensors of
the Dirac cones are
\begin{widetext}
\bea
\hat{v}_{\alpha\beta} = \left( \begin{array}{cccccccc}
    -\beta\eta t_2\cos\phi_1\sin k_y^c & &  -\beta t_2\cos\phi_1 \sin 
    k_y^c  & &  -\alpha t_{\perp}\cos\phi_2\sin \frac{k_z^c}{2} \\ 
    \beta t_2\sin\phi_1\sin k_y^c & &  \beta\eta t_2\sin\phi_1\sin 
    k_y^c & & -\alpha t_{\perp}\sin\phi_2 \sin \frac{k_z^c}{2}\\ 
    - 2 \beta\eta t_1 \sin k_y^c & & 2\beta t_1\sin k_y^c  & & 0 
\end{array}\right) .
\eea
\end{widetext}
The winding number is the sign of the determinator of the velocity
tensor $N_w(\alpha\beta)={\rm sgn}[\det(\hat{v}_{\alpha\beta})]$. One finds
\be
N_w (\alpha\beta) = - \alpha {\rm sgn}[ t_1 t_2 t_{\perp} \sin( \phi_1
+ \eta \phi_2) ] . 
\ee
Note that Weyl points in the same block $\alpha$ have the same topological
charge (independent of $\beta$), whereas Weyl points in different block have
opposite topological charge. The positions and motions of the four
Weyl points are illustrated in Fig.~2 for a specific case where
$N_w=-\alpha$, $k_y^c=-k_x^c<\pi$, and $k_z^c>\pi$. According to
Eq.~(\ref{dnc}), the Chern number $C_{xy}(k_z)$ varies with $k_z$. The
Chern number changes only when the gap is closed and re-opened, i.e.,
when $k_z$ passes through $k_z^c$ and $2\pi-k_z^c$. The Chern number
is calculated as
\be
C_{xy}= \sum_{\alpha=\pm} \frac{1}{4\pi}\int_0^{2\pi}\hspace{-0.2cm} d
k_x \int_0^{2\pi}\hspace{-0.2cm} dk_y {\bf
  n}_{\alpha}\cdot \left(\partial_{k_x}{\bf
  n}_\alpha\times \partial_{k_y}{\bf n}_\alpha\right) ,
\ee
where ${\bf n}_\alpha={\bf h}_\alpha/|{\bf h}_\alpha|$. We find that
$C_{xy}(k_z)=2{\rm sgn}[t_1t_2t_{\perp}\sin(2\phi_1)]$, when
$k_z\in(k_z^{\rm min}, k_z^{\rm max})$, otherwise $C_{xy}(k_z)=0$.
The relation between the quantum phase transitions and the evolution
of the Weyl points is depicted in Fig.~2 where we consider a
situation with $\phi_2=-0.25\pi$ and $\phi_1$ varying from $-0.1\pi$
to $0.1\pi$. At $\phi_1=-0.1\pi$, the Weyl points with $N_w<0$ are in
the $k_z<\pi$ region whereas those with $N_w>0$ are in the $k_z>\pi$
region [Fig.~2(a)]. Consequently, the Chern number $C_{xy}(k_z)$ is
negative at $k_z\in(k_z^{\rm min}, k_z^{\rm max})$. As $\phi_1$
increases the Weyl points with $N_w<0$ and those with $N_w>0$ moves in
opposite directions along $k_z$ and becomes closer. At $\phi_1=0$,
Weyl points with opposite $N_w$ {\em coincide} and the system becomes
a non-topological semimetal. After that their positions {\em shift} as
$\phi_1$ increases [Fig.~2 (c)]: the Weyl points with $N_w<0$ in the
$k_z>\pi$ region whereas those with $N_w>0$ move to the $k_z<\pi$.
As a consequence the Chern number $C_{xy}(k_z)$ changes sign at
$k_z\in (k_z^{\rm min}, k_z^{\rm max})$. Further variation of $\phi_1$ will
enlarge the region and finally the system becomes a QAH insulator
after $k_z^{\rm min}\to 0$ where pairs of monopoles with opposite
topological charge ($N_w$) merge and annihilate each other. During
those processes {\em quantized Berry fluxes are inserted into each
  $k_x$-$k_y$ plane with fixed $k_z$, whenever the monopoles move
  across it}, as there are quantized Berry fluxes flow between
monopoles with opposite charge.

In addition, as $\phi_2\to 0,\pm \pi/2$, and $\pi$, the two Dirac cones
with same $k_z$ {\em merge} together and form a quadratic band
touching as they have the same winding number. The $k_z$ dependence of
$C_{xy}(k_z)$ in this situation is similar to the previous one. However, as
we will show later, the Fermi arcs and surface states in those two
situations are significantly different.

\begin{figure}[htb]
\includegraphics[height=3.7cm]{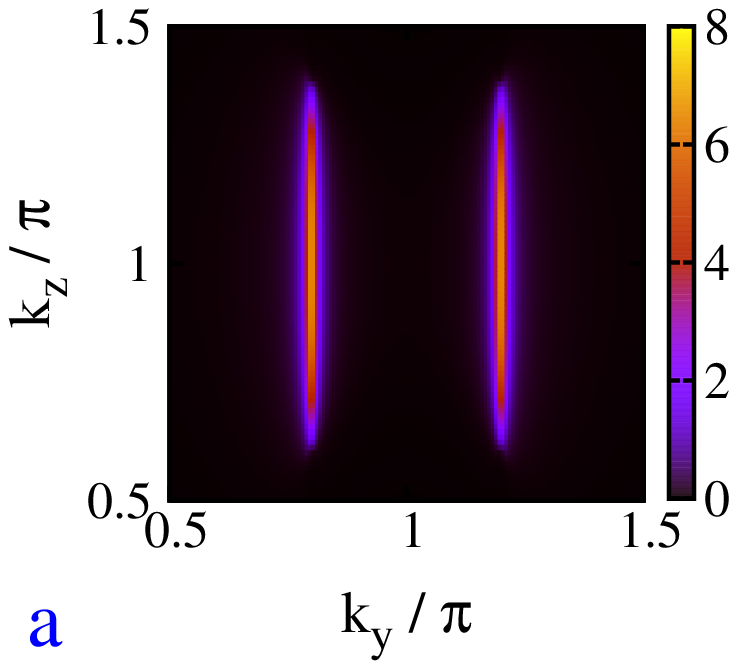}\includegraphics[height=3.7cm]{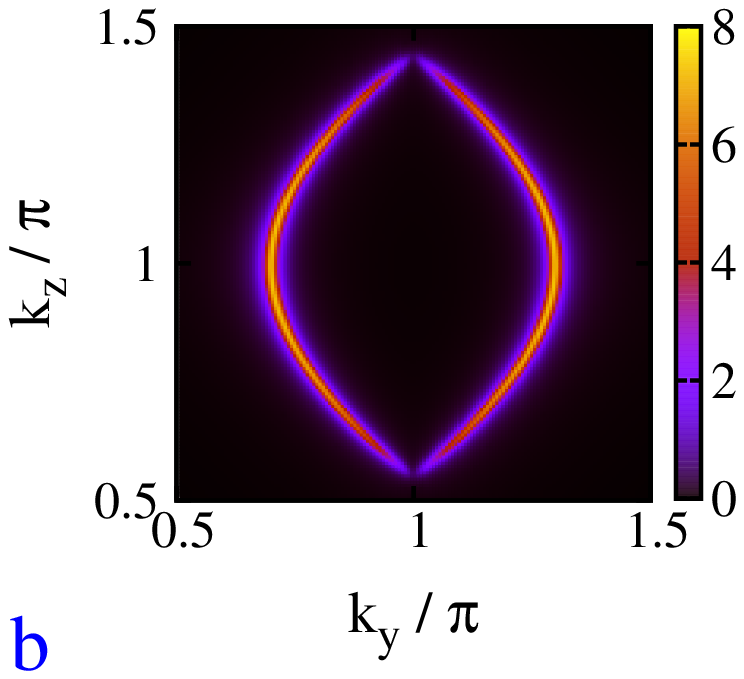}
\includegraphics[height=4.3cm]{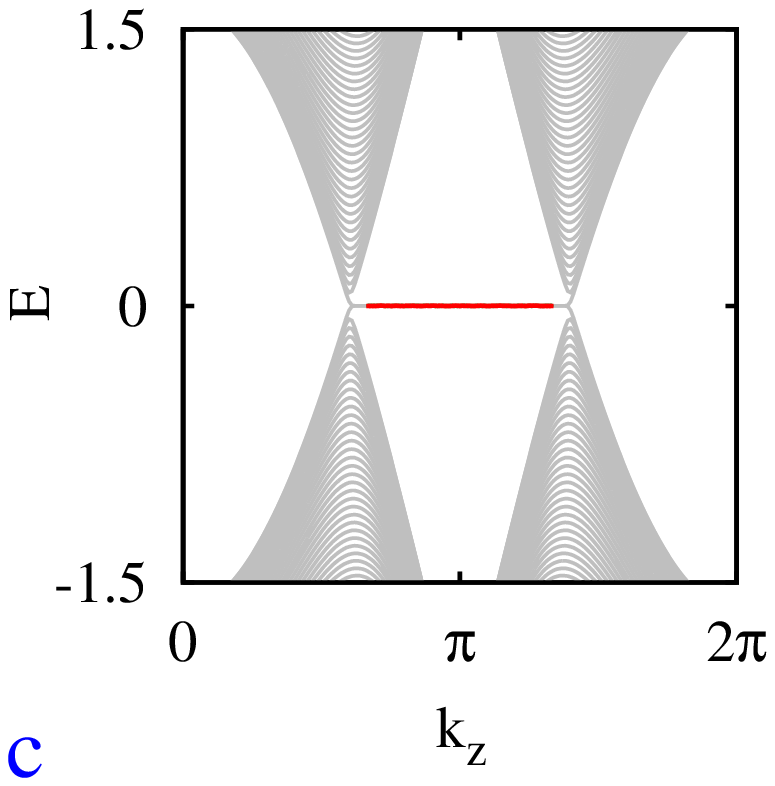}\includegraphics[height=4.3cm]{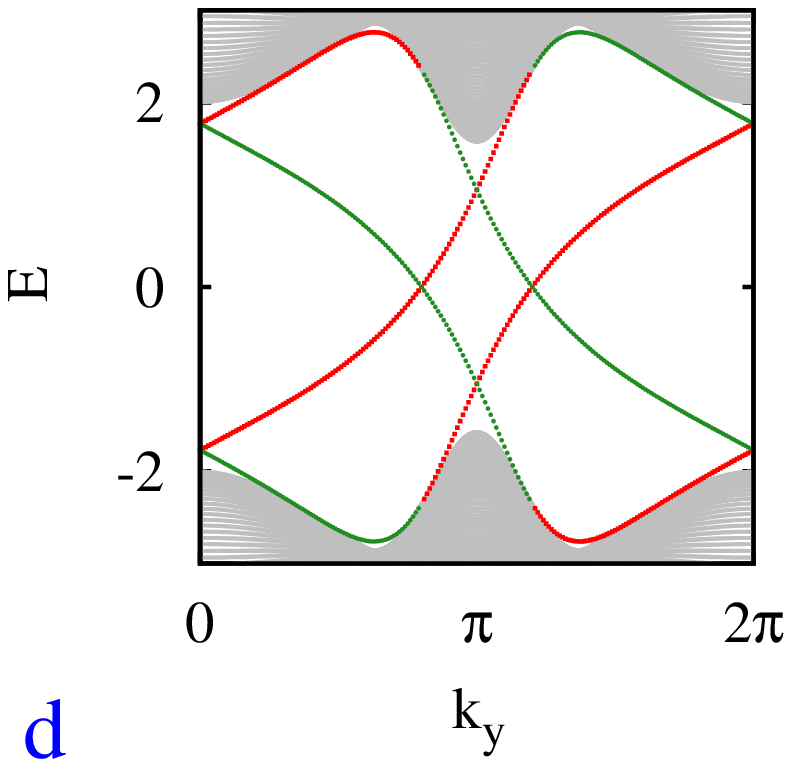}
\caption{(Color online) Spectral function ${\cal A}(E)$ of
  the surface states at zero energy ($E=0$) in the case with (a) four Dirac
  cones and (b) two quadratic band touchings. Spectra of surface and
  bulk states as function of (c) $k_z$ (at $k_y=0.8\pi$) and (d) $k_y$
  (at $k_z=\pi$) respectively with the same parameters as in (a). The
  size of the system is 151 unit cell (due to the finite size effect
  the bulk bands are slight gapped at the node). Gray region
  represents bulk spectra, while red (green) curves denote the surface
  spectra at the left (right) boundary [Note that in (c) the surface
  spectra at the two boundaries coincide]. $t_1=1$, $t_2=2$, and
  $t_{\perp}=2$. In (a), (c) and (d) $\phi_1=0.1\pi$,
  $\phi_2=-0.25\pi$, whereas in (b) $\phi_1=0.15\pi$,
  $\phi_2=-0.5\pi$. Correspondingly, $k_y^c=0.8\pi$ and $k_z^c=1.4\pi$
  in (a), (c) and (d), whereas $k_y^c=\pi$ and $k_z^c=1.5\pi$ in
  (b). The results are calculated via the iterative Green function
  method in Ref.~\cite{dai}. An artificial spectral broadening $0.04$
  is taken for the sake of visibility.} 
\end{figure}

\section{Evolution of Fermi arcs and surface states}
The move and merge (or split) of the Weyl points have profound effects
on the topologically protected surface states. In particular, we
demonstrate the Fermi arcs on the surface for two situations where
there are (i) four Dirac cones and (ii) two quadratic band touchings in
Fig.~3 (a) and (b) respectively. The color represents the spectral
function ${\cal A}(E)=\frac{-1}{\pi}{\rm Im}G^{r}(E)$ [$G^{r}(E)$ is
the retarded Green function of the system] of the zero-energy ($E=0$)
surface states (i.e., the Fermi arcs). The surface is perpendicular to
the $x$ direction. It is noted that although the $k_z$ dependence of
$C_{xy}(k_z)$ are similar in the two cases, the Fermi arcs are quite
different. This is essentially due to the different
positions of the monopoles in ${\bf k}$-space. From Fig.~3(c), one can
see that {\em the zero-energy surface states  (Fermi arcs) merge into
  the bulk bands through the Weyl points}. This figure clearly
visualizes the fact that Fermi arcs are Dirac strings which link the
monopoles and antimonopoles of opposite  ``magnetic'' charge ($N_w$)
in the Berry-phase gauge fields\cite{Volovik}. As Fermi arcs terminate at
the Weyl points, the positions of the monopoles are essential to the
shape of the Fermi arcs and the topologically protected surface states
spectra. From the figure it is seen that the monopole-antimonopole
pair with the same $k_y$ (but opposite $k_z$) are connected via the
Dirac strings. This string configuration is
due to the facts that (i) the system possesses the inversion symmetry:
$k_z\to -k_z$; (ii) the string connecting the monopoles and
anti-monopoles must pass through the $k_z=\pi$ plane because it exists
only in the region $k_z\in  (k_z^{\rm min}, k_z^{\rm max})$ dictated
by the topology. Hence the string connection between
monopole-antimonopole pair with the same $k_y$ (and opposite $k_z$) is
protected by the symmetry and dictated by the topology.

\begin{figure}[htb]
\includegraphics[height=3.7cm]{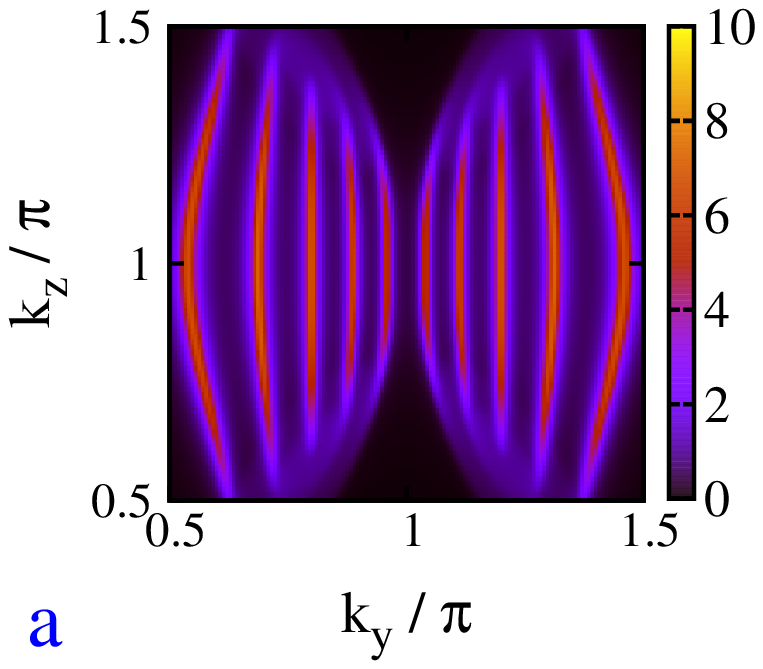}\includegraphics[height=3.7cm]{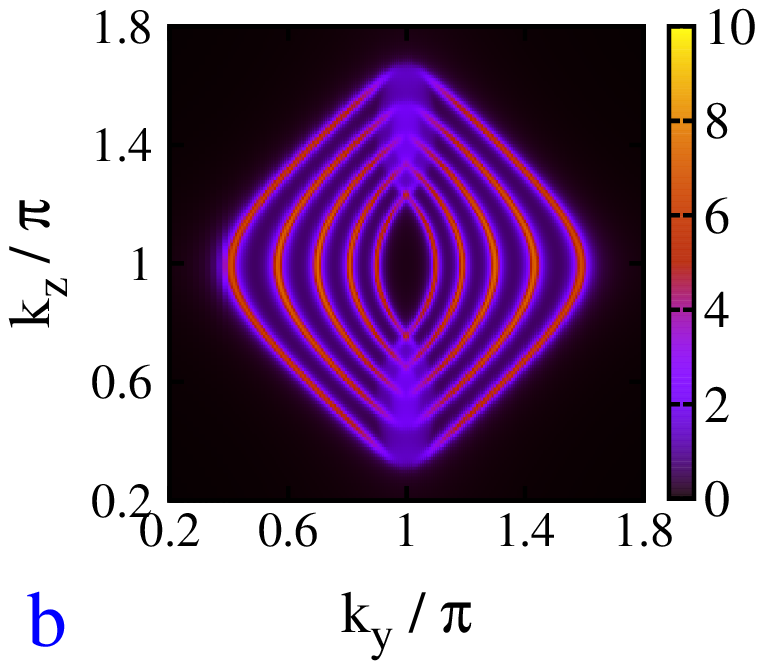}
\caption{(Color online) Spectral functions of the surface states for
  two cases: (a) four Dirac cones (b) two quadratic band touchings. The
  energies of the surface states from left to right (in both the
  $k_y<\pi$ and $k_y>\pi$ regions) are -0.8, -0.4, 0, 0.4, and 0.8
  respectively. E.g., both the leftmost arc in the $k_y<\pi$ region and
  the leftmost arc in the $k_y>\pi$ region are the arcs with energy
  -0.8. The parameters for (a) and (b) are the same as those in
  Fig.3~(a) and 3~(b) respectively. An artificial spectral broadening
  $0.04$ is taken for the sake of visibility.}
\end{figure}

It is noted from Fig.~3(b) that in case (ii) the two Fermi arcs are
actually connected together as $k_y^c=-k_y^c$, forming a closed Fermi
surface. In the four Dirac cone case in Fig.~3(a), when $k_y^c\ne 
-k_y^c$, the Fermi surface is {\em not} closed. It seems that the
Fermi arcs are also invariant under the inversion operation $k_y\to
-k_y$. However, this is only a special property of {\em zero} energy
surface states due to the particle-hole symmetry. For
surface states with {\em nonzero} energy, the spectral function are not
$k_y$-inversion symmetric as the boundary breaks the $(x,y)\to
(-x,-y)$ inversion symmetry. To illustrate the surface states
explicitly, we plot the spectral functions of the surface states in
Fig.~4(a) and 4(b) for the two cases that correspond to Fig.~3(a) and
3(b) respectively. The selected  energies of the surface states from left
to right (in both the $k_y<\pi$ and $k_y>\pi$ regions) are -0.8, -0.4,
0, 0.4, and 0.8 respectively. For each energy there are two arcs: one
in the $k_y<\pi$ region and another in the $k_y>\pi$ region. E.g.,
both the leftmost arc in the $k_y<\pi$ region and the leftmost arc in
the $k_y>\pi$ region are the arcs with energy $E=-0.8$. It is clearly
seen that the spectra are not invariant under $k_y$-inversion but
invariant under the particle-hole transformation. Besides, only the
arcs with {\em zero} energy links between the monopoles whereas other
arcs merge into the bulk bands {\em without} going through the Weyl
points. Finally, it is seen from Fig.~3(d) that the two Fermi arcs are
assigned to two different chiral edge states at the left boundary
which both have positive group velocity along the $z$-direction. It is
found that the shape of the Fermi arcs can be tuned via $\phi_1$ and
$\phi_2$ to be inner-curved (as in Fig.~3b), out-curved, or flat [as
in Fig.~3(a)]. Flat bands can be interesting in the context of
elevated transition temperature in spontaneous symmetry
broken\cite{Volovik}, as the density of states are increased.

\section{A Possible Scheme for Experimental Realization}
In this section we propose a possible scheme to realize the model in
optical lattices. The required artificial gauge fields are generated
by the spatial variation of the laser-atom interaction as suggested in
Ref.~\cite{the2}. Suppose that there is a excited states with energy
much higher than the energy scale where the above model are
defined. Impose standing waves of light to induce the following
coupling between the ground and excited states in rotating wave
approximation,
\be
H_{R}({\bf r}) = M[\cos2\pi z \cos \pi(x+y)\hat{F}_x + \cos\pi(x-y)\hat{F}_y +
\zeta \hat{F}_z] ,
\label{raman}
\ee
where $\hat{F}_\nu$ ($\nu=x,y,z$) are the Pauli matrices acting on the
(dressed-) ground and excited states. $M$ and $\zeta$ are parameters of
the laser-atom coupling. It can be realized in a system with three
standing wave Raman lasers with a detuning of $M\zeta$\cite{the2}. The
three laser wavevectors are $(\pi,\pi,\pm 2\pi)$ and $(\pi,-\pi,0)$.
When the kinetic energy is small compared to the energy splitting of
the local dressed states [the eigenstates of the local
Hamiltonian $H_{R}({\bf r})$]. The emergent gauge fields can be
obtained via the Berry-phase\cite{the2}, ${\bf A}=\bra{\Psi_G({\bf
    r})}i\grad_{\bf r}\ket{\Psi_G({\bf r})}$ with $\ket{\Psi_G({\bf
    r})}$ being the ground state of the local Hamiltonian $H_{R}({\bf
  r})$. One can show that ${\bf A} = \frac{1}{2} \Theta \grad_{\bf r}\Pi$
where $\Theta = 1-\frac{|\zeta|}{\sqrt{\cos^2(2\pi z) \cos^2
  \pi(x+y)+\cos^2\pi(x-y)+\zeta^2}}$ and $\Pi={\rm Arg}[\cos(2\pi z)
\cos\pi(x+y)+i \cos\pi(x-y)]$. The effective ``magnetic field'' (i.e.,
the Berry curvature, or the ``magnetic flux density'') are written as
$B_\nu = \varepsilon^{\nu\delta\rho} \partial_\delta A_\rho$, with
$\nu,\delta,\rho=x,y,z$ and $\varepsilon$ being the Levi-Civita
tensor. The plaquette flux for each plaquette is the ``magnetic
flux'' through it. For example, for a plaquette in the $x$-$y$ plane,
the plaquette flux is $\Phi = \int dxdy B_z$, where the integral is
limited within the plaquette. We plot the effective ``magnetic field''
$B_z$ through the $x$-$y$ plane (in odd layer) and $B_{[1\ov{1}0]}$
through the $(1\ov{1}0)$ plane in Fig.~5(a) and 5(b) respectively. It
is seen that the magnetic flux density has exactly the same
checkerboard pattern as the plaquette fluxes in Fig.~1(a) and
1(b). This shows that the gauge fields generated by the Hamiltonian,
Eq.~(\ref{raman}), is exactly what is needed for the realization of
the model. The magnetic flux density in $(110)$ plane is zero
everywhere, which is because the current scheme realizes the model
with $\phi_1=\phi_2$.

\begin{figure}[htb]
\includegraphics[height=3.7cm]{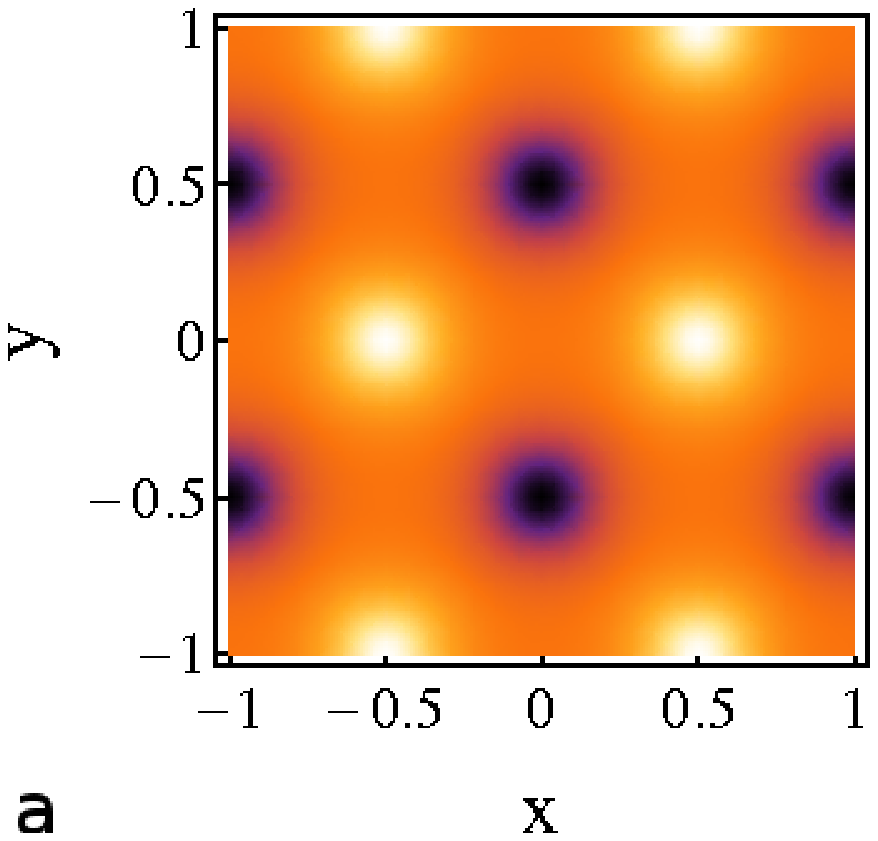}\includegraphics[height=3.7cm]{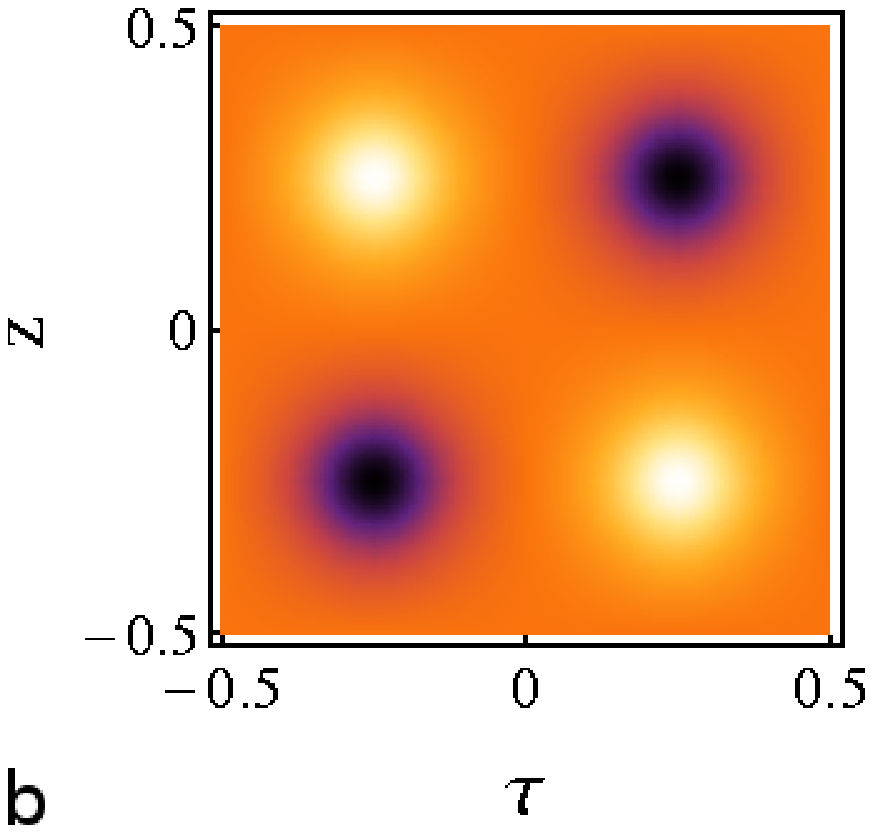}
\caption{(Color online) The ``magnetic flux density'' through (a)
  $(001)$ plane at an odd layer (b) $(1\ov{1}0)$ plane. $\tau={\bf
    r}\cdot (1,1,0)/2=(x+y)/2$. The bright region has positive value,
  whereas the dark region has negative value.}
\end{figure}

The phase along a hopping path $l$ is given by $\phi_l = \int_l
d{\bf r}\cdot {\bf A}$. Combined with Figs.~1(a) and 1(b), one can
show that the hopping phases $\phi_1$ and $\phi_2$ in Hamiltonian
Eq.~(\ref{H0}) are given by
\be
\phi_1 = \phi_2 = g(\zeta)
\ee
whereas the phases of other hoppings are zero. The function $g(\zeta)$
is plotted in Fig.~6. $\phi_1=\phi_2\in [-\pi/4,0]$. Due to the
factor $\cos(2\pi z)$ the gauge phases are inverted from odd layer at
$z=n$ to even layer at $z=n+\frac{1}{2}$ with $n$ being integer. This
exactly realizes the lattice structure in Fig.~1. The amplitudes and
signs of the rest hoppings can, in principle, be tuned via
various optical lattice techniques\cite{sgn}. By manipulating the
ratio $t_2/t_{\perp}$, the system can experience various phases in the
phase diagram: (i) At finite $\phi_1$ when $|t_2/t_{\perp}|<1$ it is a
Weyl semimetal, (ii) otherwise it is a QAH insulator with
$C_{xy}(k_z)=2{\rm sgn}[t_1t_2t_{\perp}\sin(2\phi_1)]$ for all $k_z$;
(iii) At $\phi_1=0$ the system is always a non-topological semimetal.

\begin{figure}[htb]
\includegraphics[height=3.8cm]{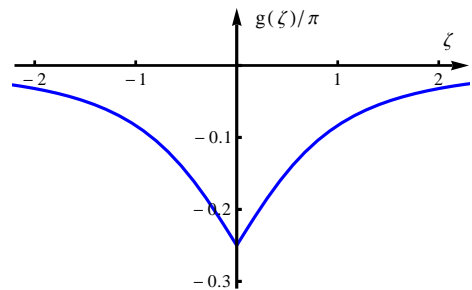}
\caption{The function $g(\zeta)$.}
\end{figure}

\section{Summary}
In summary, a model of simple cubic lattice with staggered fluxes
which exhibits Weyl semimetal phase is proposed and studied. The model
is simple and can act as a prototype to study the properties
of Weyl semimetals. Due to its simplicity, the model is potentially
achievable both in ultracold fermions in optical lattices and in
condensed matter systems. In particular, we propose a possible scheme
to realize the model in optical lattice system. Differing from
previous works, here the mechanism to achieve the topological Weyl
semimetal state is to gap the quadratic band touching by
time-reversal-symmetry-breaking hoppings. The system exhibits rich
phase diagrams, where the number of Weyl fermions and their
topological charges and positions are tunable via the plaquette fluxes
(hopping phases). The Weyl semimetal state is demonstrated to be the
intermediate phase between non-topological semimetal and quantum 
anomalous Hall insulator. The transitions between those phases can be
understood via the evolution of the Weyl points [see below]. As the
Weyl points move and split (or merge) via the manipulation of the
hopping phases, the Fermi arcs and surface states undergo significant
change, as the Fermi arcs have to be terminated at the Weyl points.

The relations between the non-topological insulators/semimetals,
topological Weyl semimetals and QAH insulators demonstrated in this
paper can be summarized in the following processes: (i)
A non-topological insulator becomes non-topological semimetal via
forming an accidental band-touching node with winding number
$N_w=0$. (ii) By {\em splitting} the node into pairs of Weyl fermions
with opposite topological charges and {\em moving} the positively and
negatively charged Weyl fermions in {\em opposite} directions in the
${\bf k}$-space, a Weyl semimetal state is created. (iii) When the
positively and negatively charged Weyl points are moved by {\em half}
of the reciprocal lattice vector, they merge and annihilate each other
in pairs, making the system transit into a QAH insulator. The {\em
  sign} of the Chern number of the QAH state depends on the {\em
  direction} along which the Weyl fermions are moved. Quantized Berry
fluxes are {\em inserted} into the bulk states during the evolution
because Weyl fermions are the monopoles of Berry-phase gauge
fields. This picture can be generalized to understand the relations
between the non-topological insulators/semimetals, the time-reversal
symmetric topological Weyl semimetals and the 3D quantum spin Hall
insulator (i.e., the time-reversal invariant $Z_2$ topological
insulators), where quantized non-Abelian Berry fluxes are inserted
into the bulk states during the moving of the monopoles.

\section*{Acknowledgments}
Work at the Weizmann Institute  was supported by the German Federal
Ministry of Education and Research (BMBF) within the framework of the
German-Israeli project cooperation (DIP) and by the Israel Science
Foundation (ISF). I thank Zhong Fang, Xi Dai, Jonathan Ruhman, and
Zohar Ringel for illuminating discussions and comments.


\begin{thebibliography}{99}
\bibitem{early} There are some early studies that are summarized in
  Ref.~\cite{Volovik}. Also Ref.~\cite{fdt} proposed that some gapless
  semiconductors with broken inversion symmetry can be candidates of
  Weyl semimetals.
\bibitem{sgn} The sign of hopping can be tuned by some subtle
  techniques. For example, in laser-assisted tunnel
  hopping\cite{the1}, the hopping phase can be tuned by the laser.
  If such phase is $\pm \pi$, then the sign of the hopping will be
  changed.
\end{thebibliography}
\end{document}